\documentclass[12pt]{iopart}

\usepackage{iopams}
\usepackage{graphicx}
\usepackage{graphics}
\usepackage{epsf}

\begin{document}

\title[Incommensurate Phase of a Triangular Frustrated Heisenberg Model ...]{Incommensurate Phase of a Triangular Frustrated Heisenberg Model Studied via Schwinger-Boson Mean-Field Theory}

\author{Peng Li}
\address{Department of Physics, Sichuan University, Chengdu 610064, China}
\ead{lipeng@scu.edu.cn}

\author{Haibin Su}
\address{Division of Materials Science, Nanyang Technological University, 50
Nanyang Avenue, Singapore 639798} \ead{hbsu@ntu.edu.sg}

\author{Hui-Ning Dong}
\address{College of Electronic Engineering, Chongqing University of Posts and
Telecommunications, Chongqing 400065, China}
\ead{donghn@cqupt.edu.cn}

\author{Shun-Qing Shen}
\address{Department of Physics, The University of Hong Kong, Pokfulam, Hong Kong, China}
\ead{sshen@hku.hk}

\begin{abstract}
We study a triangular frustrated antiferromagnetic Heisenberg
model with nearest-neighbor interaction $J_{1}$ and
third-nearest-neighbor interactions $J_{3}$ by means of
Schwinger-boson mean-field theory. It is shown that an
incommensurate phase exists in a finite region in the parameter
space for an antiferromagnetic $J_{3}$ while $J_{1}$ can be
either positive or negtaive. A detailed solution is presented to
disclose the main features of this incommensurate phase. A
gapless dispersion of quasiparticles leads to the intrinsic
$T^{2}$-law of specific heat. The local magnetization is
significantly reduced by quantum fluctuations (for $S=1$ case, a
local magnetization is estimated as $m=\left\langle
S_{i}\right\rangle \approx0.6223$). The magnetic susceptibility
is linear in temperature at low temperatures. We address possible
relevance of these results to the low-temperature properties of
NiGa$_{2}$S$_{4}$. From a careful analysis of
the incommensurate spin wave vector, the interaction parameters for NiGa$_{2}%
$S$_{4}$ are estimated as, $J_{1}\approx-3.8755$K and
$J_{3}\approx14.0628$K, in order to account for the experimental
data.

\end{abstract}


\maketitle

\section{Introduction}
In two-dimensional (2D) antiferromagnets, it was proposed the
"geometrical frustration" may enhance thequantum spin fluctuation
and suppress the magnetic order to form a spin liquid
\cite{Anderson}. In this context the triangular- and
kagom\'{e}-related lattices are studied extensively to seek
quantum spin liquid \cite{Misguich}. It turns out that the
triangular lattice antiferromagnet with nearest-neighbor (NN)
coupling exhibits $120^{\circ}$ magnetic order \cite{Huse}, while
the kagom\'{e} lattice antiferromagnet is still a controversial
topic for intriguing exploring \cite{LeePA}. People resort to
other interactions, such as longer range and multiple-spin
exchange ones, to realize quantum spin liquid \cite{Misguich}.
Experimental evidences in favor of this long-predicted
spin-liquid state have emerged in recent years \cite{Broholm},
although many aspects are still elusive. The spin disorder at low
temperatures found in\ the compound NiGa$_{2}$S$_{4}$, in which
Ni spins ($S=1$) forms a stack of triangular lattices, aroused
much attention \cite{Nakatsuji,Nakatsuji2,Nakatsuji3,Yaouanc}.
The crystal structure of the material is highly 2D, since
inter-layer interactions are quite weak. Intriguing
low-temperature properties of this material include $T^{2}$-law of
specific heat, incommensurate short-range spin correlation, and
lack of divergent behavior of the magnetic susceptibility. A
dominant third-nearest-neighbor (3rd-NN) antiferromagnetic (AFM)
interaction $J_{3}$ could produce the incommensurate phase in a
rough picture: four sublattices will form commensurate
$120^{\circ}$ magnetic order separately if the NN interaction
$J_{1}$ is zero, and the system will be driven into an
incommensurate order if $J_{1}$ is gradually switched on. A
first-principle calculation by Mazin \cite{Mazin} suggests a
large 3rd-NN interaction $J_{3}$ and a negligible 2nd-NN
interaction. $J_{3}$ is confirmed to be AFM, but the sign of
$J_{1}$ has not yet been identified \cite{Mazin}. The classical
spin version of this model was studied in a Monte-Carlo
simulation \cite{Tamura}, which provides some helpful
informations such as the incommensurability. Up to now, the
quantum spin version of this model has not yet been studied very
well. Besides the sign of $J_{1}$, many aspects of this model,
either in agreement or disagreement with the experiment of
NiGa$_{2}$S$_{4}$, need further clarification and treatments. In
this paper we focus on the low-temperature properties of the
quantum spin model and intend to make a contribution to this
topic.

The Schwinger-boson mean-field theory (SBMFT) provides a reliable
description for both quantum ordered and disordered
antiferromagnets based on the picture
of the resonant valence-bond (RVB) state \cite{Anderson,Auerbach,Auerbachbook}%
. As a merit, it does not prescribe any prior order for the
ground state in advance, which should emerge naturally if the
Schwinger bosons condense in the lowest energy states. For the
Heisenberg antiferromagnets with NN couplings at zero
temperature, it successfully captures the ($\pi,\pi$) magnetic
order on the square lattice and the $120^{\circ}$ magnetic order
on triangular lattice respectively
\cite{Auerbach,Auerbachbook,Gazza,Shen}. By means of SBMFT, we
will show that the $J_{1}$-$J_{3}$ model falls into an
incommensurate order phase at zero temperature for an AFM\
$J_{3}$ and either a FM $J_{1}$ or an AFM\ $J_{1}$ . By analyzing
the incommensurate spin wave vector, we find that the NN
interaction $J_{1}$ should in the FM region to obtain an
appropriate incommensurate phase. We also show that the
$T^{2}$-law of specific heat is an intrinsic feature of this
phase, the magnetic susceptibility is linear in temperature, and
the local magnetization is significantly reduced by quantum
fluctuations. We address possible relevance of these results to
low temperature properties of NiGa$_{2}$S$_{4}$. Our results
suggests that the
$J_{1}$-$J_{3}$ model is an essential part of the minimal model for NiGa$_{2}%
$S$_{4}$. In the following, we first present a formalism of the
SBMFT scheme for the $J_{1}$-$J_{3}$ model, then solve the
mean-field equations numerically and calculate relevant
quantities. Finally we discuss the physical meanings of the
results.

\section{The Schwinger-boson mean-field theory}
The $J_{1}$-$J_{3}$ model on the triangular lattice reads%
\begin{equation}
H=J_{1}\sum_{\left\langle ij\right\rangle \in
NN}\mathbf{S}_{i}\cdot\mathbf{S}_{j}+J_{3}\sum_{\left\langle
i^{^{\prime}}j^{^{\prime}}\right\rangle \in
3rd-NN}\mathbf{S}_{i^{^{\prime}}}\cdot\mathbf{S}
_{j^{^{\prime}}}. \label{H}
\end{equation}
We set $J_{3}>0$, but $J_{1}$ can be either AFM or FM. In the
Schwinger-boson
representation for the spin operators, $S_{i}^{+}=a_{i}^{\dag}b_{i},S_{i}%
^{-}=b_{i}^{\dag}a_{i},S_{i}^{z}=\left(  a_{i}^{\dag}a_{i}-b_{i}^{\dag}%
b_{i}\right)  /2$ with $[a_{i},a_{j}^{\dag}]=[b_{i},b_{j}^{\dag}]=\delta_{ij}%
$, we decompose the NN and 3rd-NN interactions as\cite{Li04NJP}%
\begin{equation}
J_{1}\mathbf{S}_{i}\cdot\mathbf{S}_{j}   =J_{1} :F_{ij}^{\dag}%
F_{ij}:-J_{1}A_{ij}^{\dag}A_{ij},\label{decoupleJ1}
\end{equation}
\begin{equation}
J_{3}\mathbf{S}_{i^{^{\prime}}}\cdot\mathbf{S}_{j^{^{\prime}}}   =-J_{3}%
\Pi_{i^{^{\prime}}j^{^{\prime}}}^{\dag}\Pi_{i^{^{\prime}}j^{^{\prime}}},
\end{equation}
with $F_{ij}=(a_{i}^{\dag}a_{j}+b_{i}^{\dag}b_{j})/2$, $A_{ij}=(a_{i}%
b_{j}-b_{i}a_{j})/2$, and $\Pi_{i^{^{\prime}}j^{^{\prime}}}=(a_{i^{^{\prime}}%
}b_{j^{^{\prime}}}-b_{i^{^{\prime}}}a_{j^{^{\prime}}})/2$.
Correspondingly, we introduce three competing mean fields,
$F=\left\langle F_{ij}\right\rangle $, $A=-i\left\langle
A_{ij}\right\rangle $, and $\Pi=-i\left\langle
\Pi_{i^{^{\prime}}j^{^{\prime}}}\right\rangle $, and apply the
Hartree-Fock decompositions for the interactions. A Lagrangian
multiplier $\lambda$ is also introduced to impose the constraint
on the Schwinger bosons, $+\lambda\sum _{i}\left(
a_{i}^{\dag}a_{i}+b_{i}^{\dag}b_{i}-2S\right)  $. After performing
the Fourier's transform, the effective Hamiltonian can be written
in a compact
form,%
\begin{equation}
H_{eff}=\sum_{\mathbf{k}}\phi_{\mathbf{k}}^{\dag}M\left(
\mathbf{k}\right) \phi_{\mathbf{k}}+\varepsilon_{0},
\end{equation}
where $\phi_{\mathbf{k}}^{\dag}
=(a_{\mathbf{k}}^{\dag},b_{\mathbf{k}}^{\dag
},a_{-\mathbf{k}},b_{-\mathbf{k}}), M\left(  \mathbf{k}\right)
=\epsilon\left( \mathbf{k}\right)  \sigma
_{0}\otimes\sigma_{0}+\Delta\left( \mathbf{k}\right)
\sigma_{y}\otimes \sigma_{y}, \epsilon\left( \mathbf{k}\right)
=\lambda-J_{1}F\sum_{\delta}\cos k^{(\delta)}, \Delta\left(
\mathbf{k}\right)   =J_{1}A\sum_{\delta}\sin k^{(\delta
)}+J_{3}\Pi\sum_{\delta}\sin2k^{(\delta)}, \varepsilon_{0}
=3N_{\Lambda}(-J_{1}F^{2}+J_{1}A^{2}+J_{3}\Pi
^{2})-N_{\Lambda}\lambda\left(  2S+1\right)$, and $\otimes$ means
the Kronecker product, $\sigma_{0}$ is a $2\times2$ unit matrix,
$\sigma_{\alpha}$'s $(\alpha=x,y,z)$ are Pauli matrices,
$k^{(\delta
)}=k_{x},k_{x}/2+\sqrt{3}k_{y}/2,-k_{x}/2+\sqrt{3}k_{y}/2$ for
$\delta=1,2,3$ respectively. The Matsubara Green's function are
defined as,
\begin{equation}
G\left(  \mathbf{k},\tau\right)  =-\left\langle T_{\tau}\phi_{\mathbf{k}%
}\left(  \tau\right)  \phi_{\mathbf{k}}^{\dag}\right\rangle ,
\end{equation}
where $\tau$ is the imaginary time and $\phi_{\mathbf{k}}\left(
\tau\right) =e^{\tau H_{eff}}\phi_{\mathbf{k}}e^{-\tau H_{eff}}$.
All physical quantities can be expressed in terms of the matrix
elements of the Green's function.

The Matsubara Green's function in Matsubara frequency
$\omega_{n}=2n\pi/\beta$ ($n$ is an integer for bosons) can be
worked out as
\begin{equation}
G(\mathbf{k},i\omega_{n}) =\frac{i\omega_{n}\sigma_{z}\otimes\sigma_{0}-\epsilon\left(
\mathbf{k}\right)  \sigma_{0}\otimes\sigma_{0}+\Delta\left(  \mathbf{k}%
\right)  \sigma_{y}\otimes\sigma_{y}}{\left(  i\omega_{n}\right)  ^{2}%
-\omega^{2}\left(  \mathbf{k}\right)  }.
\end{equation}
From the poles of the Matsubara Green's function, the two
degenerate spectra of the quasi-particles can be readily read out,
\begin{equation}
\omega\left(  \mathbf{k}\right)  =\sqrt{\epsilon^{2}\left(
\mathbf{k}\right)
-\Delta^{2}\left(  \mathbf{k}\right)  }. \label{spectrum}%
\end{equation}
The mean-field equations are established by the constraint and
the introduced mean fields. We omit the details and only present
the results here,
\numparts
\begin{eqnarray}
\frac{1}{N_{\Lambda}}\sum_{\mathbf{k}}\left(  1+2n_{B}\left[
\omega\left( \mathbf{k}\right)  \right]  \right)
\frac{\epsilon\left(  \mathbf{k}\right)
}{\omega\left(  \mathbf{k}\right)  }  =2S+1,\label{MFEa}\\
\frac{1}{6N_{\Lambda}}\sum_{\mathbf{k}}\left(  1+2n_{B}\left[
\omega\left( \mathbf{k}\right)  \right]  \right)
\frac{\epsilon\left(  \mathbf{k}\right) \sum_{\delta}\cos
k^{(\delta)}}{\omega\left(  \mathbf{k}\right)  }
=F,\label{MFEb}\\
\frac{1}{6N_{\Lambda}}\sum_{\mathbf{k}}\left(  1+2n_{B}\left[
\omega\left( \mathbf{k}\right)  \right]  \right)
\frac{\Delta\left(  \mathbf{k}\right) \sum_{\delta}\sin
k^{(\delta)}}{\omega\left(  \mathbf{k}\right)  }
=A,\label{MFEc}\\
\frac{1}{6N_{\Lambda}}\sum_{\mathbf{k}}\left(  1+2n_{B}\left[
\omega\left( \mathbf{k}\right)  \right]  \right)
\frac{\Delta\left(  \mathbf{k}\right)
\sum_{\delta}\sin2k^{(\delta)}}{\omega\left(  \mathbf{k}\right)
}   =\Pi,
\label{MFEd}%
\end{eqnarray}
\endnumparts
where $n_{B}\left[  \omega\left(  \mathbf{k}\right)  \right]
=\left[ e^{\omega\left(  \mathbf{k}\right)  /k_{B}T}-1\right]
^{-1}$ is the Bose-Einstein distribution function. In the
thermodynamical limit $N_{\Lambda }\rightarrow\infty$, the
momentum sum is replaced by an integral,
$(1/N_{\Lambda})\sum_{\mathbf{k}}\rightarrow(1/A_{BZ})\int
d^{2}k$, $A_{BZ}=8\pi^{2}/\sqrt{3}$. If the Schwinger bosons
condensation occurs at $\mathbf{k}^{\ast}$, a condensation term
should be extracted in the momentum summation of the first
equation, Eq. (\ref{MFEa}),
\begin{equation}
2S+1=\rho_{0}+\int\frac{d^{2}k}{A_{BZ}}\left(  1+2n_{B}\left[
\omega\left( \mathbf{k}\right)  \right]  \right)
\frac{\epsilon\left(  \mathbf{k}\right) }{\omega\left(
\mathbf{k}\right)  },
\end{equation}
where the density of condensates
\begin{equation}
\rho_{0}=\frac{1}{N_{\Lambda}}\sum_{\mathbf{k}^{\ast}}\left(
1+2n_{B}\left[ \omega\left(  \mathbf{k}^{\ast}\right)  \right]
\right)  \frac{\epsilon \left(  \mathbf{k}^{\ast}\right)
}{\omega\left(  \mathbf{k}^{\ast}\right)  }.
\label{rho0}%
\end{equation}
Our numerical solution demonstrates the condensation occurs at
zero temperature for spin $S>S_{C}$ with $S_{C}\lesssim0.172$.
Thus we will count condensations in later discussions of this
paper. The condensation terms in the next three mean-field
equations, Eq. (\ref{MFEb})-(\ref{MFEd}), should also be
extracted carefully. It is noticeable the per site ground state
energy can be simplified by utilizing the mean-field equations,
\begin{equation}
E_{0}/N_{\Lambda}=\frac{1}{N_{\Lambda}}\left(
\sum_{\mathbf{k}}\omega\left(
\mathbf{k}\right)  +\varepsilon_{0}\right)  =-3J_{1}(A^{2}-F^{2})-3J_{3}%
\Pi^{2}%
\end{equation}

\section{The incommensurate phase solution}
The mean-field equations are solved numerically at zero
temperature. For our purpose, we set $S=1$ in the calculation in
order to compare the result with the related experiment, although
the qualitative conclusion is spin-independent, but the
quantitative results vary with the values of spin. One fact that
should be noticed is that the mean fields $F$ and $A$ could not
exist simultaneously \cite{Gazza,Gazza2}, so the number of
mean-field equations can be reduced from $4$ to $3$ in both
$J_{1}>0$ and $J_{1}<0$ regions. In the two regions, we found the
system falls into the incommensurate
phases with gapless excitations.%

\begin{figure}
[ptb]
\begin{center}
\includegraphics{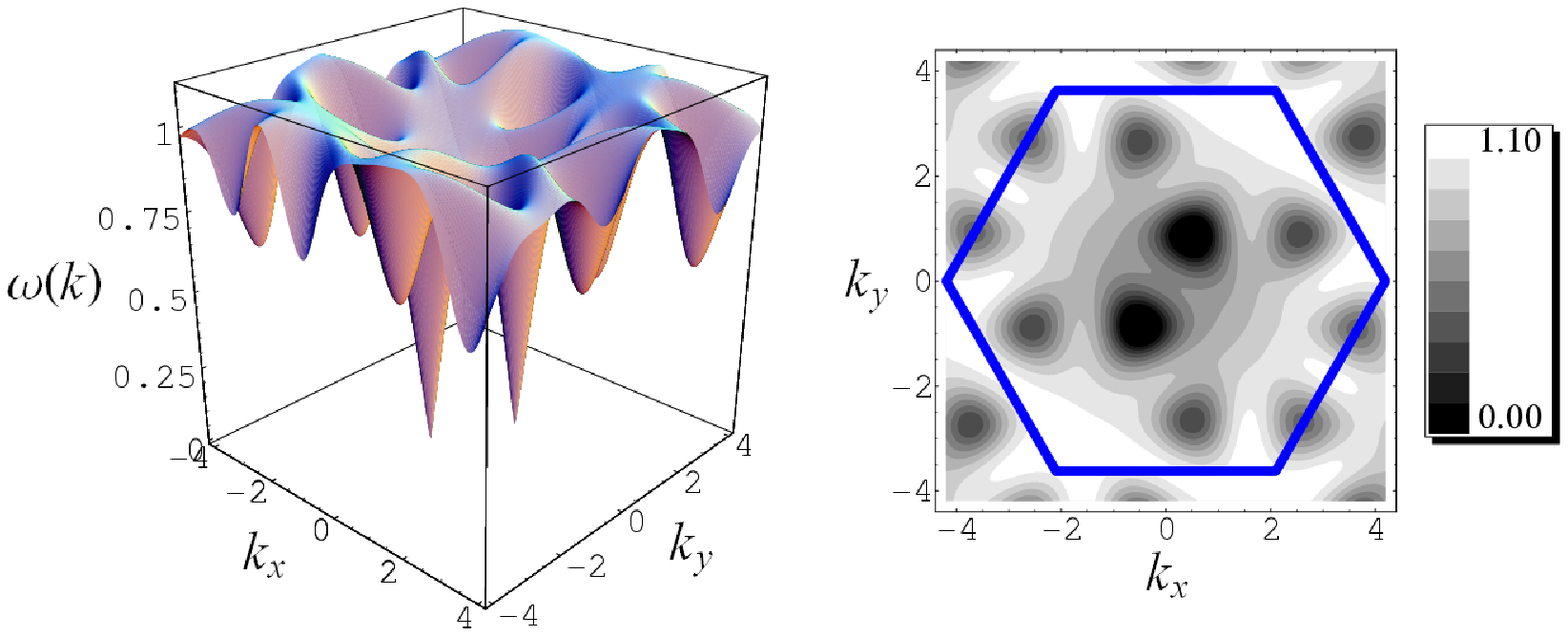}
\caption{(Color online) The gapless spectrum with nodal points.
To compare with the experiment, we choose parameter
$J_{1}/J_{3}=-0.2756$, so that the gapless nodal points occur at
$\mathbf{k}^{\ast}=\pm(k^{\ast}/2,\sqrt {3}k^{\ast}/2)$ with
$k^{\ast}=0.158\pi$. The blue hexagon denotes the first
Brillouin zone. See more details in the text.}%
\end{center}
\end{figure}

The quasiparticle's spectra become gapless at the nodal points,
say
$\mathbf{k}^{\ast}=(k_{x}^{\ast},k_{y}^{\ast})=\pm(k^{\ast}/2,\sqrt{3}k^{\ast
}/2)$ (e.g. see Fig. 1). Near the nodal points, the spectrum is
linear in $\left\vert \mathbf{k}%
-\mathbf{k}^{\ast}\right\vert,$
\begin{equation}
\omega\left(  \mathbf{k}\right)  \approx\alpha\left\vert \mathbf{k}%
-\mathbf{k}^{\ast}\right\vert +O\left(  \left\vert \mathbf{k}-\mathbf{k}%
^{\ast}\right\vert ^{2}\right)  .\label{linearW}%
\end{equation}
At a finite temperature, a gapful spectrum will develop
asymptotically as $\Delta_{gap}=c_{1}e^{-c_{2}/T}$ with constants
$c_{1}$ and $c_{2}$, which coincides with the Mermin-Wagner
theorem \cite{Auerbachbook}. The incommensurate order at zero
temperature of the system is signalled by the
divergence in the static spin structure factor,%
\begin{equation}
\chi_{S^{z}}\left(  \mathbf{q}\right)  =\frac{1}{N_{\Lambda}}\sum_{\mathbf{k}%
}\frac{1}{2}\left[  P\left(  \mathbf{k}+\mathbf{q}\right)  Q\left(
\mathbf{k}\right)  -R\left(  \mathbf{k}+\mathbf{q}\right)  R\left(
\mathbf{k}\right)  \right]  ,\label{XSz}%
\end{equation}
where $P\left(  \mathbf{k}\right)  =\left[  \epsilon\left(
\mathbf{k}\right) /\omega\left(  \mathbf{k}\right)  +1\right]
/2$,$Q\left(  \mathbf{k}\right) =\left[  \epsilon\left(
\mathbf{k}\right)  /\omega\left(  \mathbf{k}\right) -1\right]
/2$,$R\left(  \mathbf{k}\right)  =\Delta\left(  \mathbf{k}\right)
/\left[  2\omega\left(  \mathbf{k}\right)  \right]  $. Because
the spectra is gapless at $\mathbf{k}^{\ast},$ $\omega\left(
\mathbf{k}^{\ast}\right)  =0$,
$\chi_{S^{z}}\left(  \mathbf{q}\right)  $ becomes divergent at $\mathbf{q}%
^{\ast}=2\mathbf{k}^{\ast}$ (see Fig. 2),%

\begin{figure}
[ptb]
\begin{center}
\includegraphics{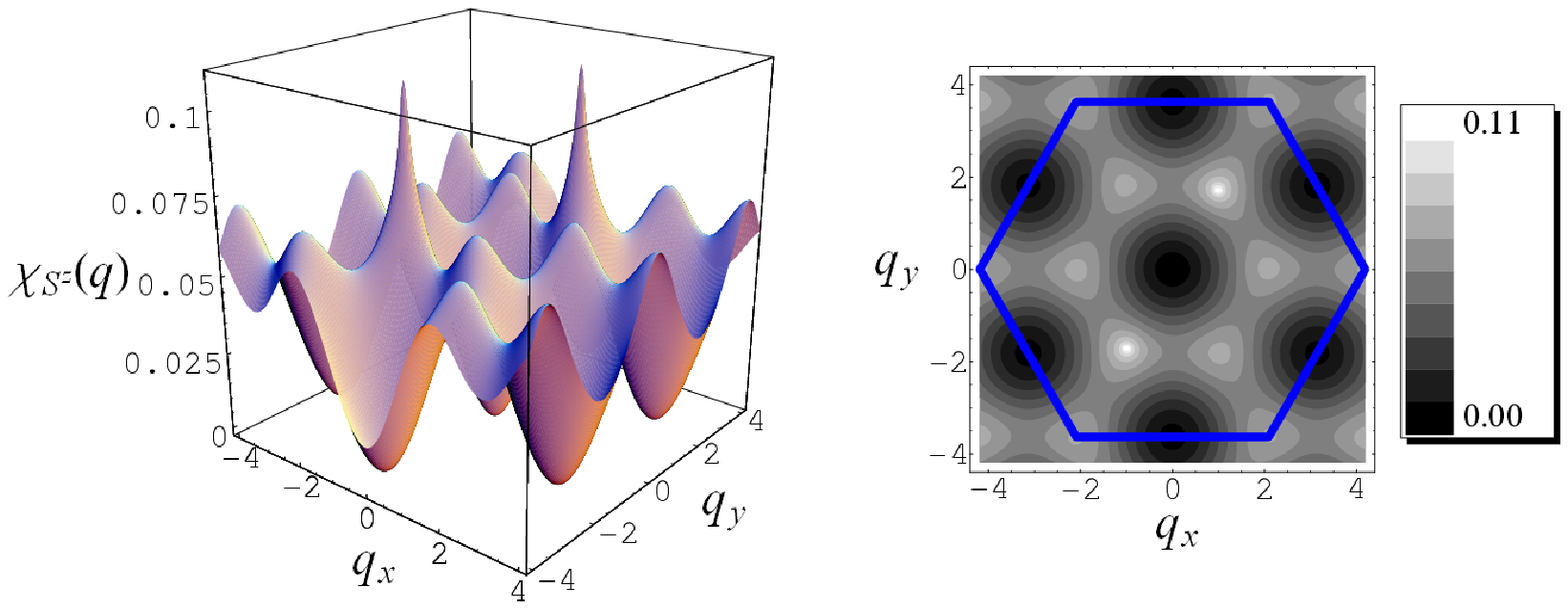}
\caption{(Color online) The zero-temperature static
spin structure factor at the parameter $J_{1}/J_{3}=-0.2756$. The
blue hexagon denotes the first
Brillouin zone. The divergent peaks located at $\mathbf{q}^{\ast}%
=2\mathbf{k}^{\ast}$ indicate an incommensurate order.}%
\end{center}
\end{figure}

\begin{equation}
\chi_{S^{z}}\left(  \mathbf{q}^{\ast}\right)
=\frac{1}{16}N_{\Lambda}\rho
_{0}^{2},\label{divergent}%
\end{equation}
as it is proportional to the number of lattice sites
$N_{\Lambda}$. The local magnetization will be reduced
significantly due to strong quantum
fluctuations,%
\begin{equation}
m\approx\sqrt{\frac{\chi_{S^{z}}\left(  \mathbf{q}^{\ast}\right)  }%
{N_{\Lambda}\left\vert \cos\mathbf{q}^{\ast}\right\vert }}=\frac{\rho_{0}%
}{4\sqrt{\left\vert \cos\mathbf{q}^{\ast}\right\vert }}.\label{m}%
\end{equation}
The important difference between the regions of $J_{1}>0$ and
$J_{1}<0$ is the nodal point's momentum $k^{\ast}\in\left[
\pi/6,\pi/3\right]  $ for $J_{1}>0$ and $k^{\ast}\in\left[
0,\pi/6\right]  $ for $J_{1}<0$ regions, respectively. In the
limit of $J_{1}/J_{3}\rightarrow\infty$,
$k^{\ast}\rightarrow\pi/3$, the solution reproduces $120^{\circ}$
spin order correctly. While below the critical value
$J_{1}/J_{3}\approx-3.71$, the system becomes a saturated
ferromagnet, where the linear expansion, Eq. (\ref{linearW}),
will be replaced by a parabolic form $\omega\left(
\mathbf{k}\right)  \approx\beta\left(
\mathbf{k}-\mathbf{k}^{\ast}\right)  ^{2}$. The plots of
$k^{\ast}$ versus $J_{1}/J_{3}$ and $\alpha$ versus $J_{1}/J_{3}$
are shown in Fig. 3.

\begin{figure}
[ptbptb]
\begin{center}
\includegraphics{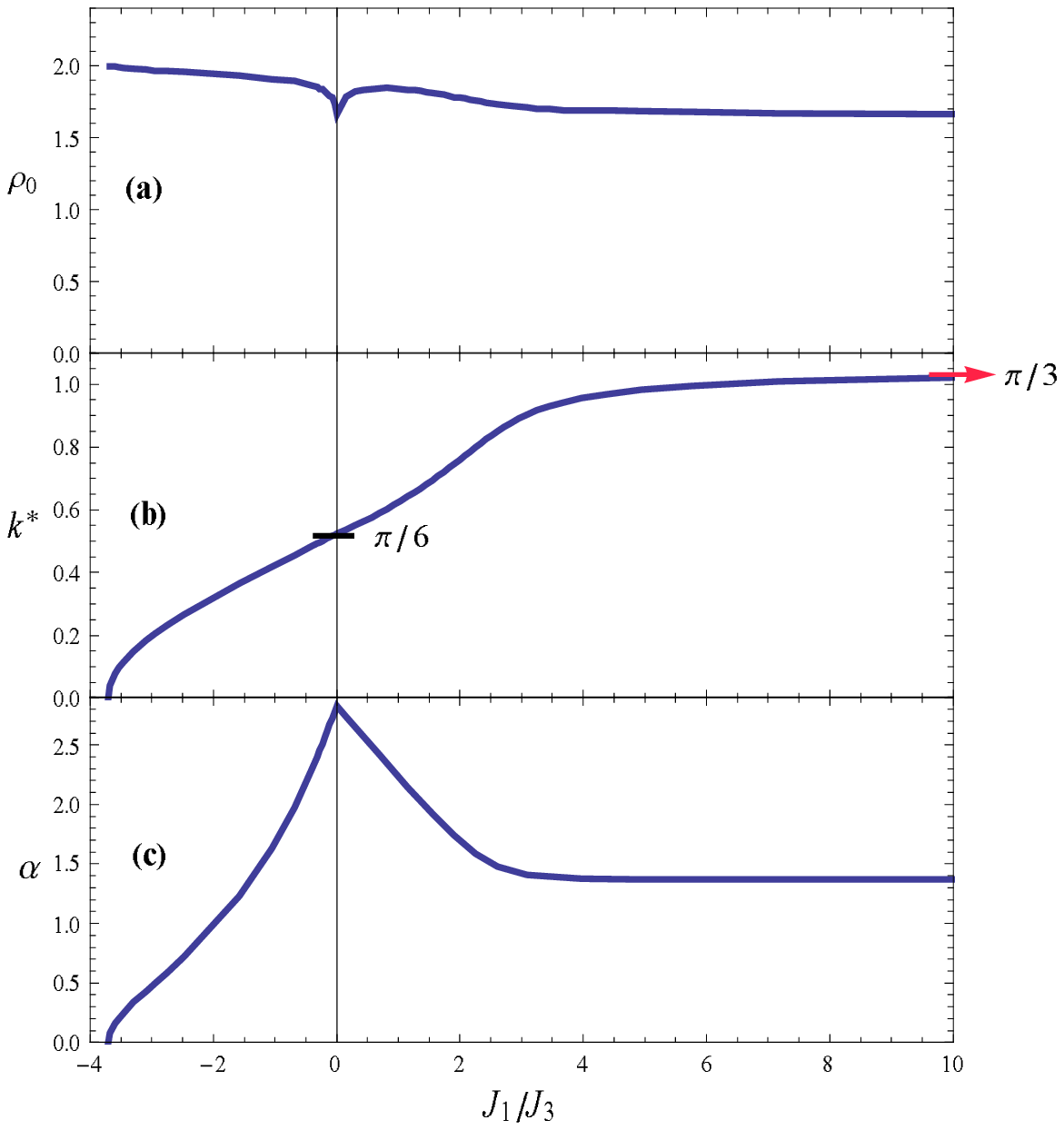}
\caption{(Color online) (a) The condensation term $\rho_{0}$
versus $J_{1}/J_{3}$. (b) The magnitude of the nodal point's
momentum of the spectrum $k^{\ast}$ versus $J_{1}/J_{3}$. In the
limit $J_{1}/J_{3}\rightarrow+\infty$, the result reproduces the
$120^{\circ}$ commensurate spin order correctly. The
incommensurate spin wave vector observed in NiGa$_{2}$S$_{4}$, $k^{\ast}%
\cong0.158\pi$, lies in the $J_{1}<0$ region. Please see more
details in the
text. (c) The coefficient $\alpha$ in Eq. (\ref{linearW}) versus $J_{1}/J_{3}%
$.}%
\end{center}
\end{figure}

The incommensurate spin wave vector observed in NiGa$_{2}$S$_{4}$
is $k^{\ast
}\cong0.158\pi<\pi/6$. From this data we estimate that $J_{1}/J_{3}%
\approx-0.2756$ from Fig. 3, which is slight different from the
value $-0.20$ in Ref.\cite{Nakatsuji}, \textit{i.e.} we have a
considerable FM $J_{1}$. Thus we can exclude the possibility of
AFM $J_{1}$\ \cite{Mazin}. The local magnetization at this point
evaluated by Eq. (\ref{m}) is $0.6223$, (not $S=1$), while the
experimental data of NiGa$_{2}$S$_{4}$ suggest a larger value,
$0.75(8)$ \cite{Nakatsuji}.

The nodal structure of the spectra, Eq. (\ref{linearW}), leads to
a linear density-of-states (DOS) in energy $E$,
\begin{equation}
D\left(  E\right)  =2\sum_{\mathbf{k}}\delta\left(  E-\omega\left(
\mathbf{k}\right)  \right)
\approx\frac{\sqrt{3}}{\pi\alpha^{2}}E.
\end{equation}
where the factor $2$ comes from the degeneracy of the
quasiparticle spectra.
As a result, a $T^{2}$-law of specific heat follows apparently,%
\begin{equation}
C_{V}/N_{\Lambda}  \approx\frac{6\sqrt{3}\zeta\left(  3\right)  k_{B}^{3}}{\pi\alpha^{2}%
J_{3}^{2}}T^{2}, \label{CV}%
\end{equation}
where $\zeta\left(  3\right)  =1.202$. If one supposes that the
$T^{2}$-law of specific heat of NiGa$_{2}$S$_{4}$ is ascribed to
the gapless incommensurate
phase, a numerical estimation, $J_{1}\approx-3.8755$K and $J_{3}%
\approx14.0628$K, could be obtained. This result is reasonable
compared to the experimental estimation $J_{3}\cong30$K
\cite{Tamura1}.

\section{Discussions}

Before ending this paper, we point out that the zero-field
susceptibility for
this incommensurate phase is linear in temperature,%
\begin{equation}
\chi_{M}/N_{\Lambda}\approx\frac{\sqrt{3}\left(  g\mu_{B}\right)  ^{2}k_{B}%
}{2\pi\alpha^{2}J_{3}^{2}}T.
\end{equation}
Using the parameters noted above, we find that it is
$\chi_{M}\approx 2.77\times10^{-4}T$(emu/mole), which is not in
agreenment with the experimental data of NiGa$_{2}$S$_{4}$,
$\chi_{M}\approx A+BT$ with $A\approx0.009$(emu/mole) and
$B\approx0$ below $10$K \cite{Nakatsuji}. The Monte-Carlo study
also shows the classical version of this model only produce a
single peak in the specific heat \cite{Tamura}. These facts
indicate that
the model in Eq. (\ref{H}) may not account for all mysteries in NiGa$_{2}%
$S$_{4}$. Thus, the solution shows the model Eq. (\ref{H}) with
AFM $J_{3}$ and FM $J_{1}$ has captured the main features for an
incommensurate correlation in NiGa$_{2}$S$_{4}$, but it is still
oversimplified as the minimal model for all low temperature
properties of NiGa$_{2}$S$_{4}$. A biquadratic interaction might
be a good candidate for reproducing a finite susceptibility at
zero temperature. In the absence of the 3rd-NN interactions, a
biquadratic term can induce a quadrupolar order and totally
suppress\ the spin order. The $T^{2}$-law of specific heat is
also intact when quadrupolar order sets in
\cite{Tsunetsugu,Lauchli,Li}. It will be interesting to see how
the incommensurate spin correlation be influenced by the
biquadratic interactions.

\section{Acknowledgement}

This work was supported by the COE-SUG Grant (No. M58070001) of
NTU and the Research Grant Council of Hong Kong under Grant No.:
HKU 703804.


\begin{thebibliography}{10}
\bibitem {Gazza2}Introduction of both $F$ and $A$ mean fields can be found in
Ref. \cite{Gazza}. Here we find that if one suppose both $F$ and
$A$ are nonzero at the same time, then an inconsistent result
will be induced. This fact is very useful when one solves the
equations. The merit of retaining both mean fields is that a
quite good ground-state energy value can be produced.
\end{thebibliography}
\end{document}